\documentclass[12pt]{iopart}

\usepackage{graphicx,hyperref,amssymb}

\begin{document}

\title{Dynamic propensity in a kinetically constrained lattice gas}

\author{Lester O. Hedges and Juan P. Garrahan}

\address{School of Physics and Astronomy, University of Nottingham,
Nottingham, NG7 2RD, UK}

\begin{abstract}
We apply the concept of dynamic propensity to a simple kinetically
constrained model of glass formers, the two-vacancy assisted
triangular lattice gas, or (2)-TLG.  We find that the propensity
field, defined in our case as the local root-mean square displacement
averaged over the ensemble of trajectories with identical initial
configurations, is a good measure of dynamical heterogeneity.  This
suggests a configurational origin for spatial fluctuations of the
dynamics, but just as in the case of atomistic systems, we find that
propensity is not correlated to any simple structural property.  We
show instead that certain extended clusters of particles connected to
vacancies correlate well with propensity, indicating that these are
the fundamental excitations of the (2)-TLG.  We also discuss
time-correlations and the correlation between configurations within
the propensity ensemble.
\end{abstract}

\date{\today}

\section{Introduction}

On approach to their glass transition \cite{ReviewsGT} glass forming
systems display increasingly heterogeneous dynamics \cite{ReviewsDH}.
This dynamic heterogeneity is not correlated in any obvious way to
structural features.  In order to uncover a possible configurational
origin for dynamic heterogeneity Harrowell and coworkers recently
proposed the concept of dynamic propensity \cite{Harrowell,Harrowell2}: a
particle's propensity is defined as some measure of its mobility over
a period of time, such as its mean square displacement, averaged over
all dynamic trajectories which start from the same initial
configuration.  This ensemble of trajectories is sometimes referred to
as the iso-configurational ensemble.  Dynamic propensity was found to
be a good indicator of dynamic heterogeneity in simulations of
atomistic models \cite{Harrowell} for times at least as long as the
structural relaxation time.  While this suggested a configurational
origin for the heterogeneity in the dynamics, no correlation was found
between dynamic propensity and simple structural measures such as
local free volume or local composition \cite{Harrowell}.

In this paper we apply the idea of dynamic propensity to a kinetically
constrained model (KCM) of a glass former, the two-vacancy assisted
triangular lattice gas \cite{Jackle}, or (2)-TLG.  We show that the
heterogeneity in the dynamics can be described by an appropriately
defined propensity field.  We find that propensity is not correlated
to simple local configurational properties such as local density.
Instead we show that a non-local structural property, the
connectedness of particles to clusters of vacancies, serves as a good
predictor of dynamic propensity.  We also discuss the behaviour of
cross-correlations between configurations in the iso-configurational
ensemble and its relation to the dynamic phase separation associated
with dynamic heterogeneity.

\begin{figure}[t]
\begin{center}
  \includegraphics[width=0.2\columnwidth]{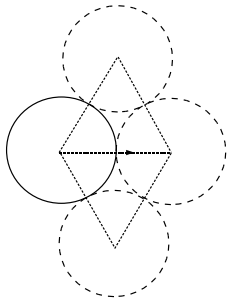}\\
  \caption{The kinetic constraint for nearest neighbour particle jumps
  in the two-vacancy assisted triangular lattice gas. Dashed lines
  refer to unoccupied sites.}\label{tlg-schematic}
\end{center}
\end{figure}

\section{(2)-TLG model and propensity}

\begin{figure}[b]
  \centerline{\hbox{ \hspace{0.0in}
    \includegraphics[width=0.33\columnwidth]{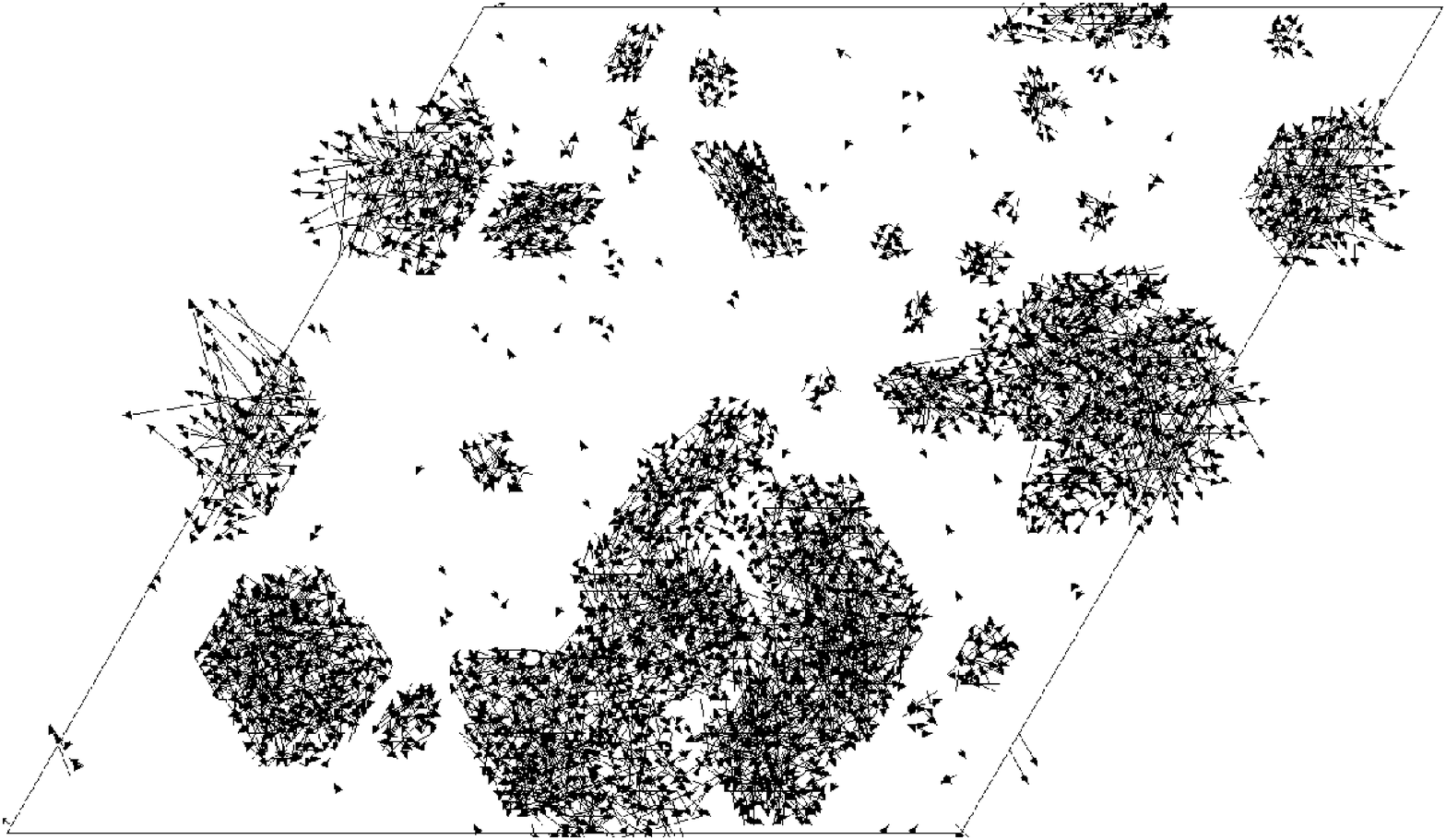}\\
    \includegraphics[width=0.33\columnwidth]{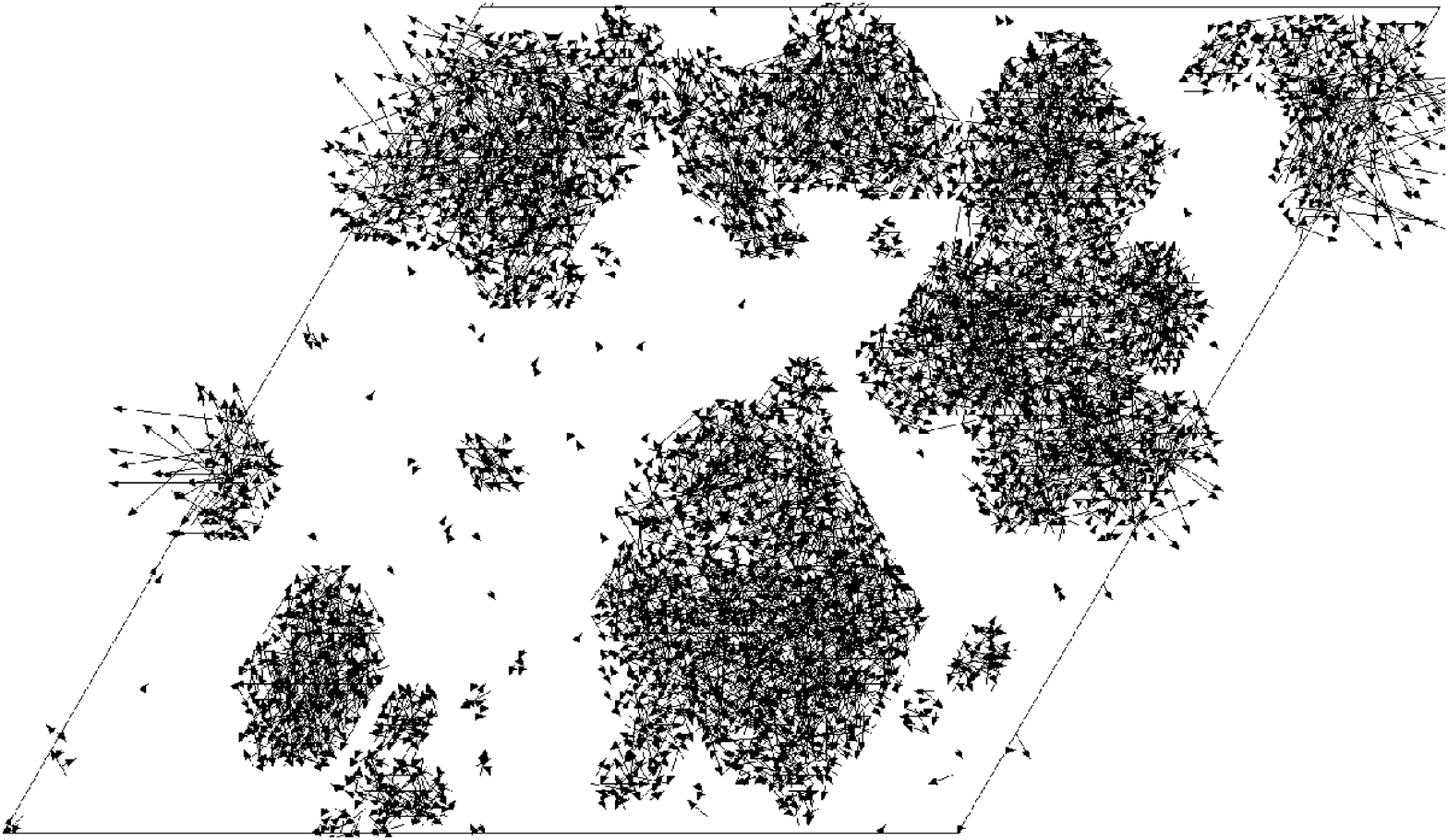}\\
    \includegraphics[width=0.33\columnwidth]{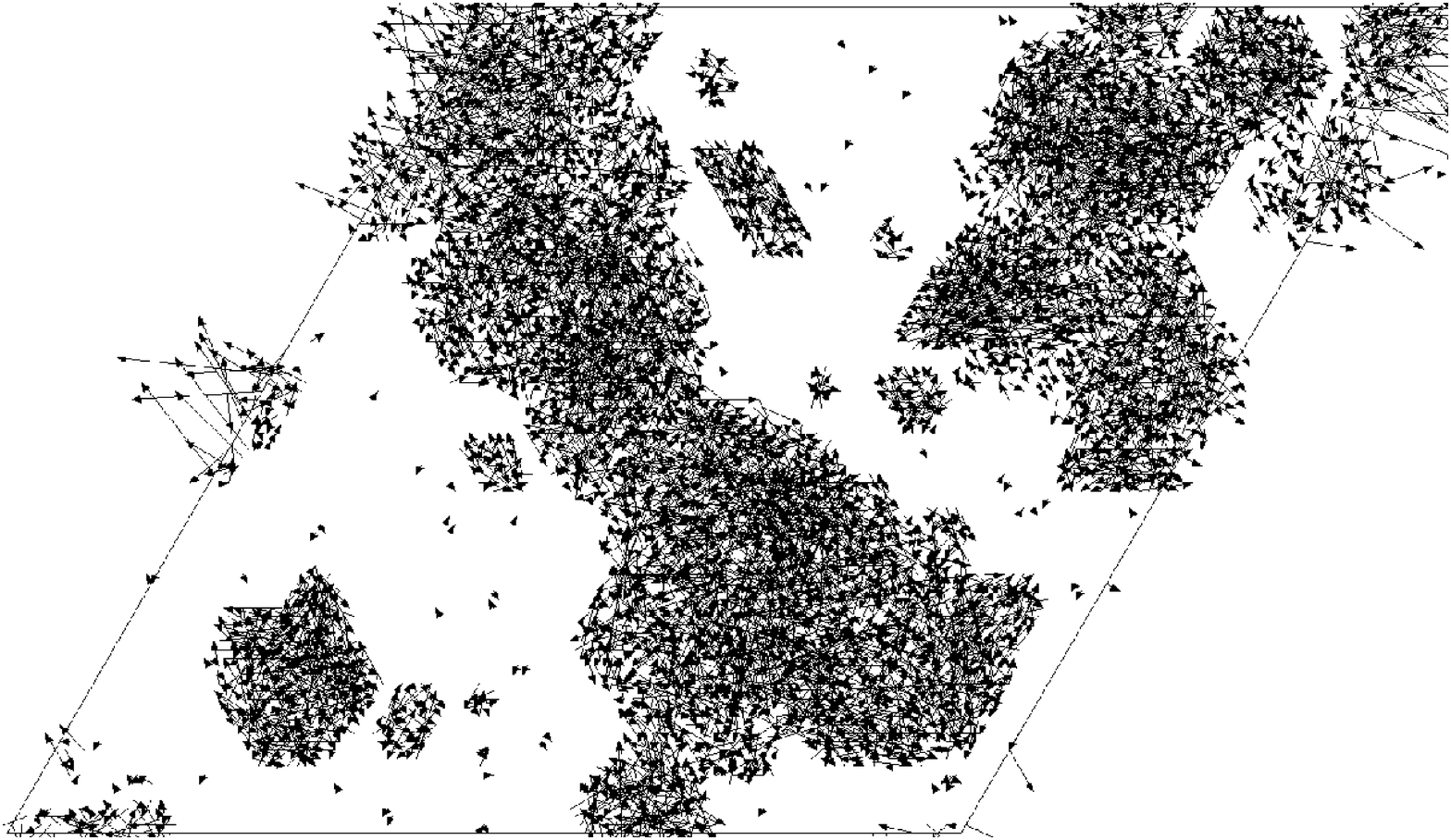}\\
    }
  }
\caption{Three particle trajectories from the same equilibrium
configuration at $\rho=0.79$.  Particle displacements are shown as
arrows joining the initial and final positions.  Each trajectory is
$t=10^6$ Monte Carlo sweeps in length and differs only in the random
sequence of attempted particle moves.} \label{trajectories}
\end{figure}

We consider the lattice gas model introduced by J\"{a}ckle and
Kr\"{o}nig \cite{Jackle}.  This model is itself a variant of the
constrained lattice models proposed by Kob and Andersen
\cite{Kob-Andersen,KA,TBF,RitSol03}.  It consists of a set of
particles that move on a two-dimensional lattice of triangular
geometry. There are no static correlations between particles and at
most each site can only hold one particle at a time. Any particle on
the lattice can only move to one of its six nearest neighbour sites if
the following rules are satisfied: (i) the target site is unoccupied
and (ii) both the two mutual nearest neighbours of the initial and
target site are also empty, see Fig.\ \ref{tlg-schematic}.  The model
is often referred to as that of two-vacancy assisted hopping on the
triangular lattice, or more simply the (2)-TLG. The absence of any
static correlations allows initial configurations to be constructed by
randomly placing particles into the lattice until the desired density
is reached. Although not intended to represent a physical system the
dynamical rules can be interpreted as the steric constraint on
particle motion within a dense fluid. For increasing particle density
the model shows a rapid dynamical slowdown and clear dynamical
heterogeneity \cite{Pan}. The dynamics of the model are indicative of
a fragile glass former \cite{Pan}.

\begin{figure}[t]
  \centerline{\hbox{ \hspace{0.0in}
    \includegraphics[width=0.5\columnwidth]{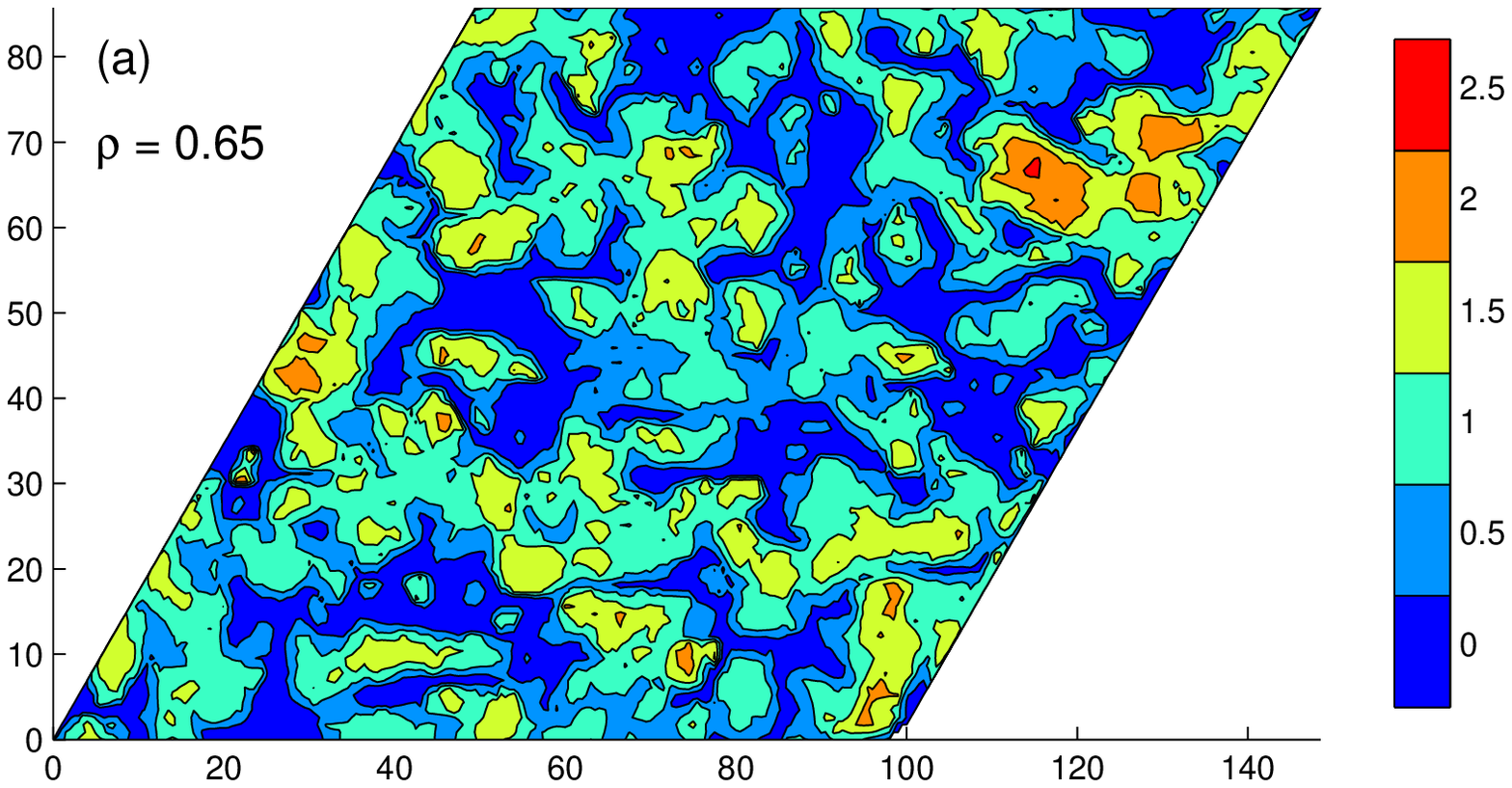}\\
    \includegraphics[width=0.5\columnwidth]{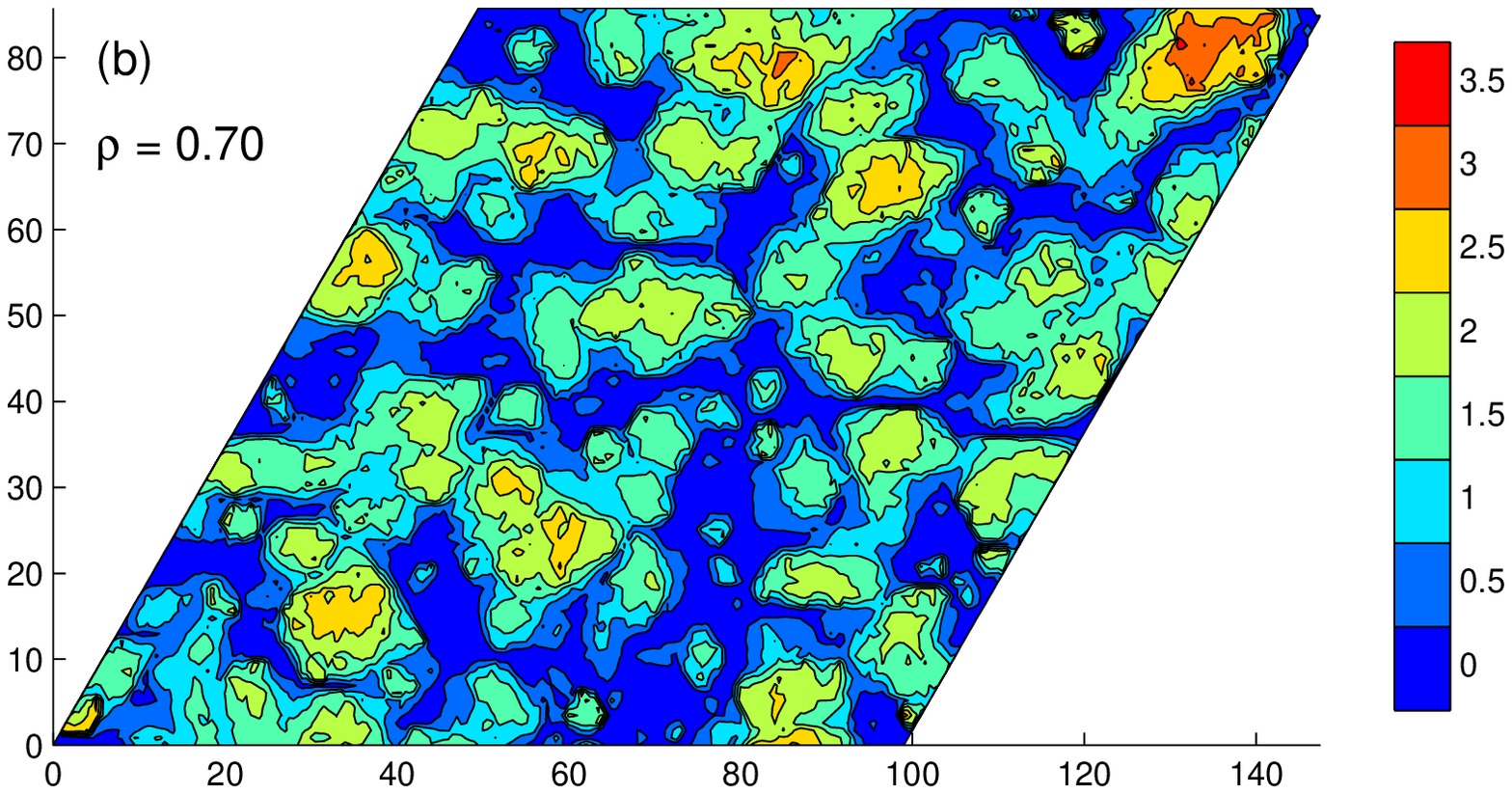}\\
    }
  }
  \vspace{9pt}
  \centerline{\hbox{ \hspace{0.0in}
    \includegraphics[width=0.5\columnwidth]{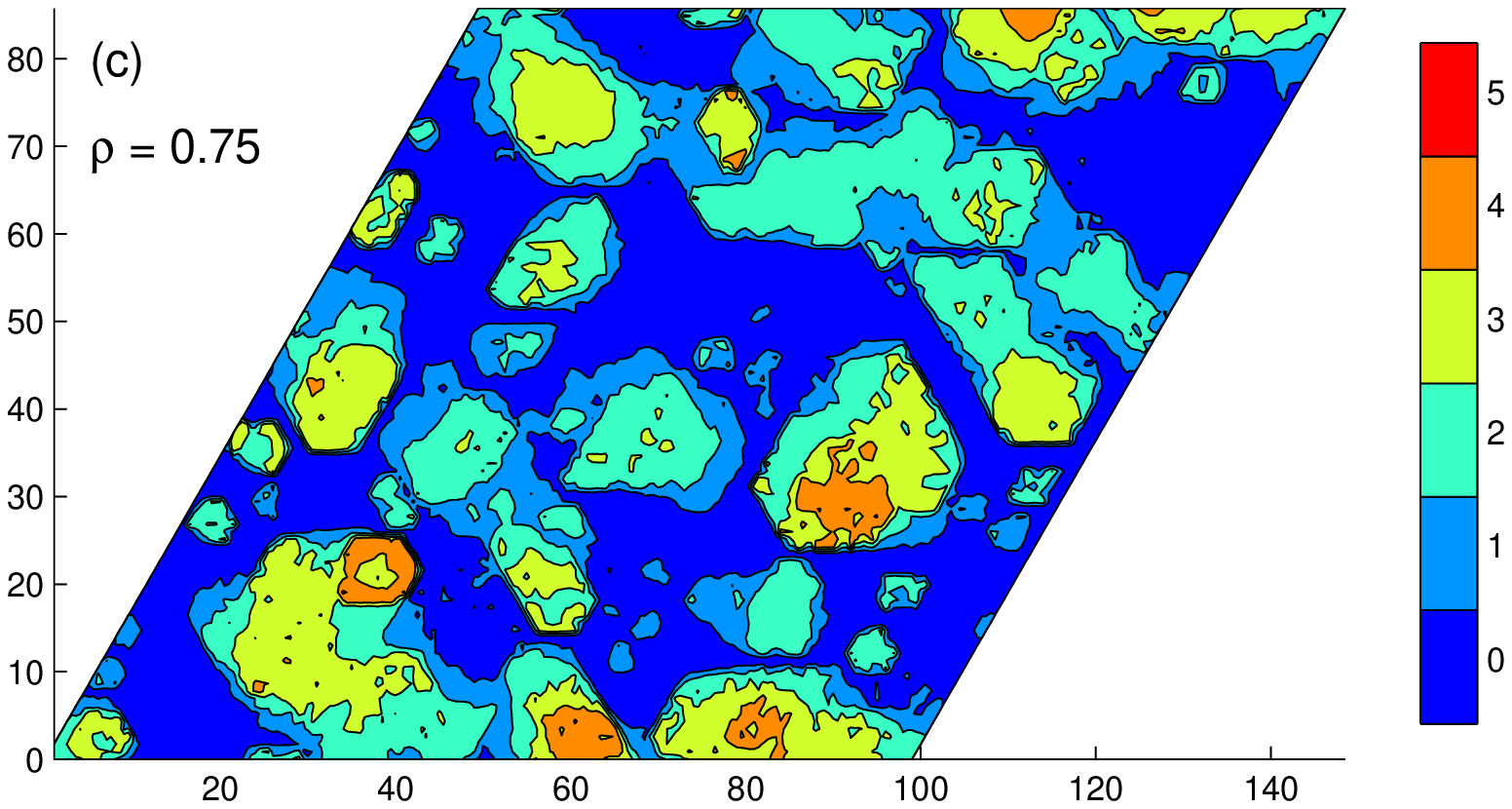}\\
    \includegraphics[width=0.5\columnwidth]{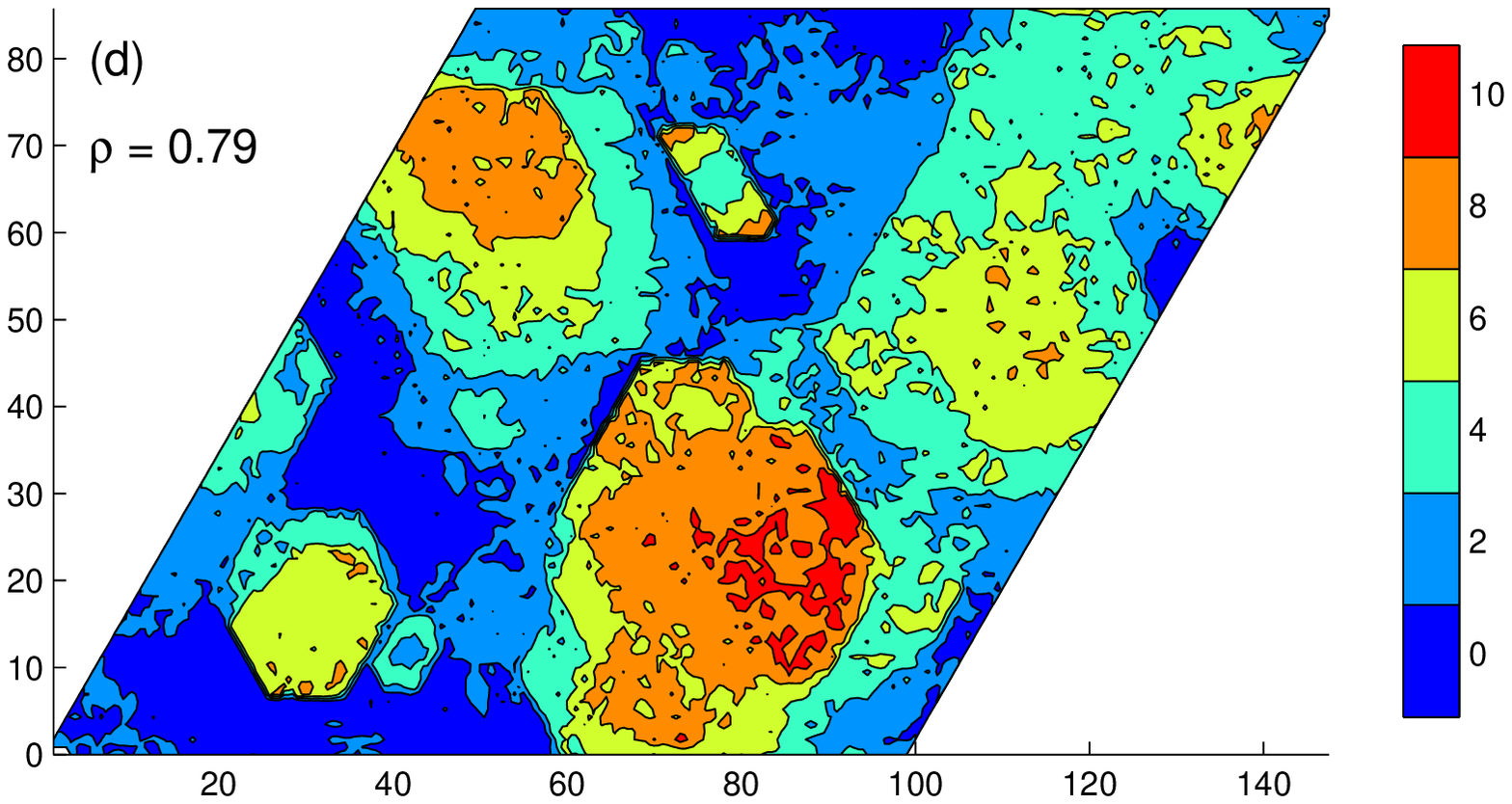}\\
    }
  }
\caption{Propensity maps for the (2)-TLG at four different particle
densities, $\rho=0.65$ (a), $\rho=0.7$ (b), $\rho=0.75$ (c) and
$\rho=0.79$ (d). Each map has been created by averaging over 100
independent trajectories.}\label{propensity-maps}
\end{figure}

For our model we choose to define the propensity of particle $i$ as
its root-mean squared displacement, $\sqrt{\langle \Delta {r_i}^2
\rangle}$.  With this choice propensity has units of length (rather
than length squared as in \cite{Harrowell}).  The average is over all
trajectories starting from the same initial configuration using
lattice gas Monte Carlo (MC) dynamics (or continuous time MC for high
densities \cite{Newman}).  That is, for a given initial condition our
propensity ensemble is that of all possible randomly attempted
particle moves.  This ensures that the dynamics of the system has
opportunity to proceed via a different route during each run: 
Fig.\ \ref{trajectories} shows the resulting particle displacements for
three different trajectories starting from the same $\rho=0.79$
initial configuration.  The propensity ensemble as defined here is the
analog in the MC case to that of randomized momenta used in MD
simulations \cite{Harrowell}.

Fig.\ \ref{propensity-maps} shows the spatial distribution of
propensity at four different particle densities $\rho$.  The times for
each density is the relaxation time $\tau_{\alpha}(\rho)$ extracted
from the persistence function \cite{Pan}.  The data is represented as
a contour plot.  Each propensity map has been averaged over $100$
independent trajectories with a lattice size of $100 \times 100$
sites. The relaxation time at $\rho = 0.79$ is approximately $10^4$
times larger than of the lowest density shown, $\rho=0.65$.  A
comparison between the four panels shows the distribution of
propensity becoming increasingly spatially heterogeneous as the
density is increased.  High propensity regions become more localised
in nature whilst domains of low propensity are observed to grow in
size.  This is similar to what is observed in atomistic systems as
temperature is decreased \cite{Harrowell}.

\begin{figure}[t]
    \centering
    \includegraphics[angle=-90,width=0.7\columnwidth]{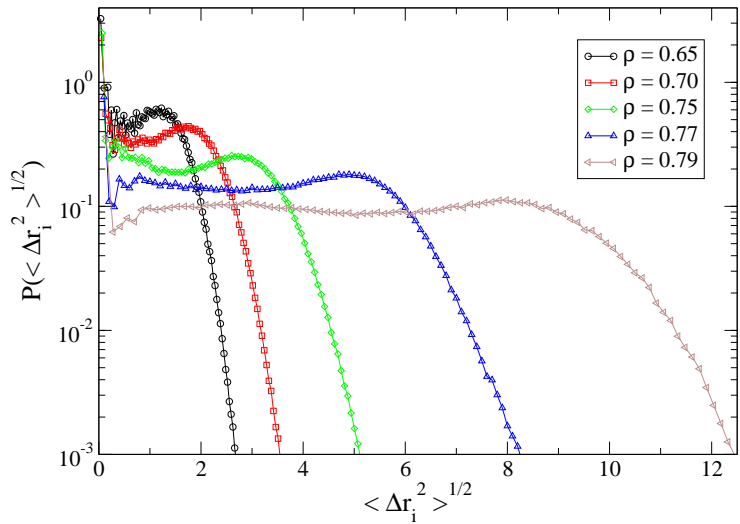}\\
    \caption{Distribution of propensities at various particle
    densities $\rho$ and times
    $\tau_\alpha(\rho)$.}\label{propensity-distribution}
\end{figure}

It is also informative to analyse the distribution of particle
propensities as a function of density, Fig.\
\ref{propensity-distribution}.  This is analogous to the van Hove
function. The figure shows that the distribution widens significantly
for increasing particle density.  Interestingly, at high density the
distributions appear remarkably flat over a large range of
propensities.  This suggests that for these time scales, the curve is
not well fit by two Gaussian profiles (or a delta and a Gaussian), one
describing the dynamics of low propensity particles, the other
characteristic of high propensity behaviour.  As argued in
Ref. \cite{EPL}, this is a sensible approximation at times either much
smaller or much larger than $\tau_\alpha$.  For the times comparable
to $\tau_\alpha$ shown here the overall distribution appears to be
that of a convolution of Gaussian functions corresponding to particles
which began diffusing at all times between 0 and $t$ \cite{EPL}.  This
seems to be an important feature of transport decoupling which
deserves further exploration.

\section{Linking structure and dynamics}

We next try to ascertain what aspect of the initial structure, if any,
is responsible for the heterogeneity observed in the spatial
distribution of propensity.  As a starting point it is useful to first
consider a simple local property of the system such as the local
density.  In any instantaneous configuration the ability of a particle
in the (2)-TLG to make a single step is entirely determined by its
local neighbourhood as this governs the dynamic rules of the
system. Those particles which have a higher number of vacancies
amongst their nearest neighbours will be more likely to move. If local
density does correlate well then one would expect the local
neighbourhood of high propensity particles to be markedly different
from those for which the propensity is low.

\begin{figure}[t]
    \centering
    \includegraphics[angle=-90,width=0.7\columnwidth]{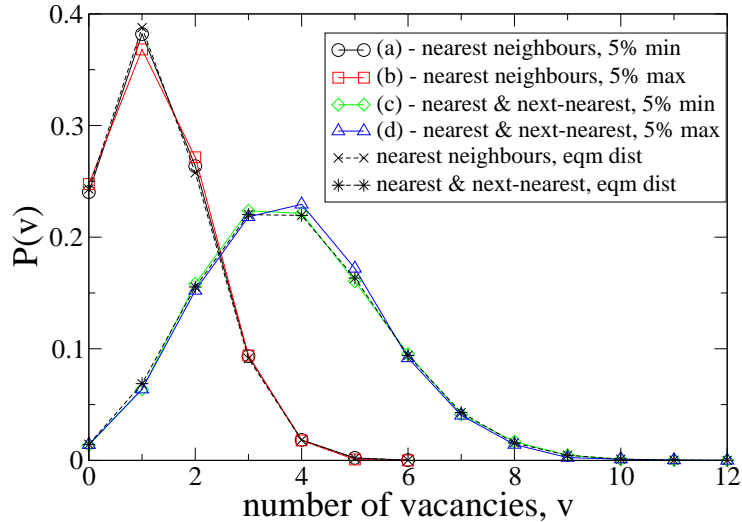}\\
    \caption{Distribution of vacancies amongst, nearest neighbours for
particles within the lowest 5\% propensity (a), nearest neighbours for
particles within the highest 5\% propensity (b), nearest and
next-nearest neighbours for particles within the lowest 5\% propensity
(c), and nearest and next-nearest neighbours for particles within the
highest 5\% propensity (d).  The figure shows that these distributions
coincide with those of a random environment at this
density.}\label{vacancy-distribution}
\end{figure}

Fig.\ \ref{vacancy-distribution} shows the distribution of vacancies
amongst nearest neighbours for particles within the lowest 5\%
propensity, nearest neighbours for particles within the highest 5\%
propensity, nearest and next-nearest neighbours for particles within
the lowest 5\% propensity, nearest and next-nearest neighbours for
particles within the highest 5\% propensity. It is clear that there is
nothing special about the local environment of the highest propensity
particles, as the distribution of vacancies appears to exactly match
the low propensity curve, even when coarse-grained over several
particle distances.  This is again analogous to what was found in
atomistic simulations \cite{Harrowell}.  In a system like the (2)-TLG,
for which the dynamics are highly collective, many cooperative
rearrangements are required in order for a particle to undergo a
significant displacement.  As such it is perhaps unsurprising that
local density fails to provide a suitable prediction of the long time
propensity for motion. In order to provide a more accurate prediction
of the spatial distribution one needs to consider some non-local
feature of the initial structure.  Interestingly, in the less
cooperative (1)-TLG \cite{Pan}, where dynamics proceeds through
diffusion of vacancy pairs, we find a similar lack of correlation
between propensity and structure as that given by Fig.\
\ref{vacancy-distribution}.

It is not individual vacancies but rather clusters of vacancies that
help facilitate the rearrangement of particles in some collective
manner \cite{KA,RitSol03,Pan}.  Considering this aspect of the
dynamics we choose to classify particles in the initial configuration
according to the size of vacancy clusters to which they are
immediately connected.  For example, although a particle may only have
a single vacancy amongst its nearest neighbours this vacancy could
itself be part of a cluster of appreciable size. A feature of this
nature would not be picked up by any localised measure.  Using a 
Hoshen-Kopelman algorithm \cite{Hoshen} we identify and
label all distinct vacancy clusters within a given initial
configuration. Following the labeling procedure it is trivial to
determine the size of each cluster. Each particle is then assigned a
value equal to the sum of the vacancy cluster sizes to which it is
connected. We define this value as the cluster connectivity.

\begin{figure}[t]
  \centerline{\hbox{ \hspace{0.0in}
    \includegraphics[width=0.5\columnwidth]{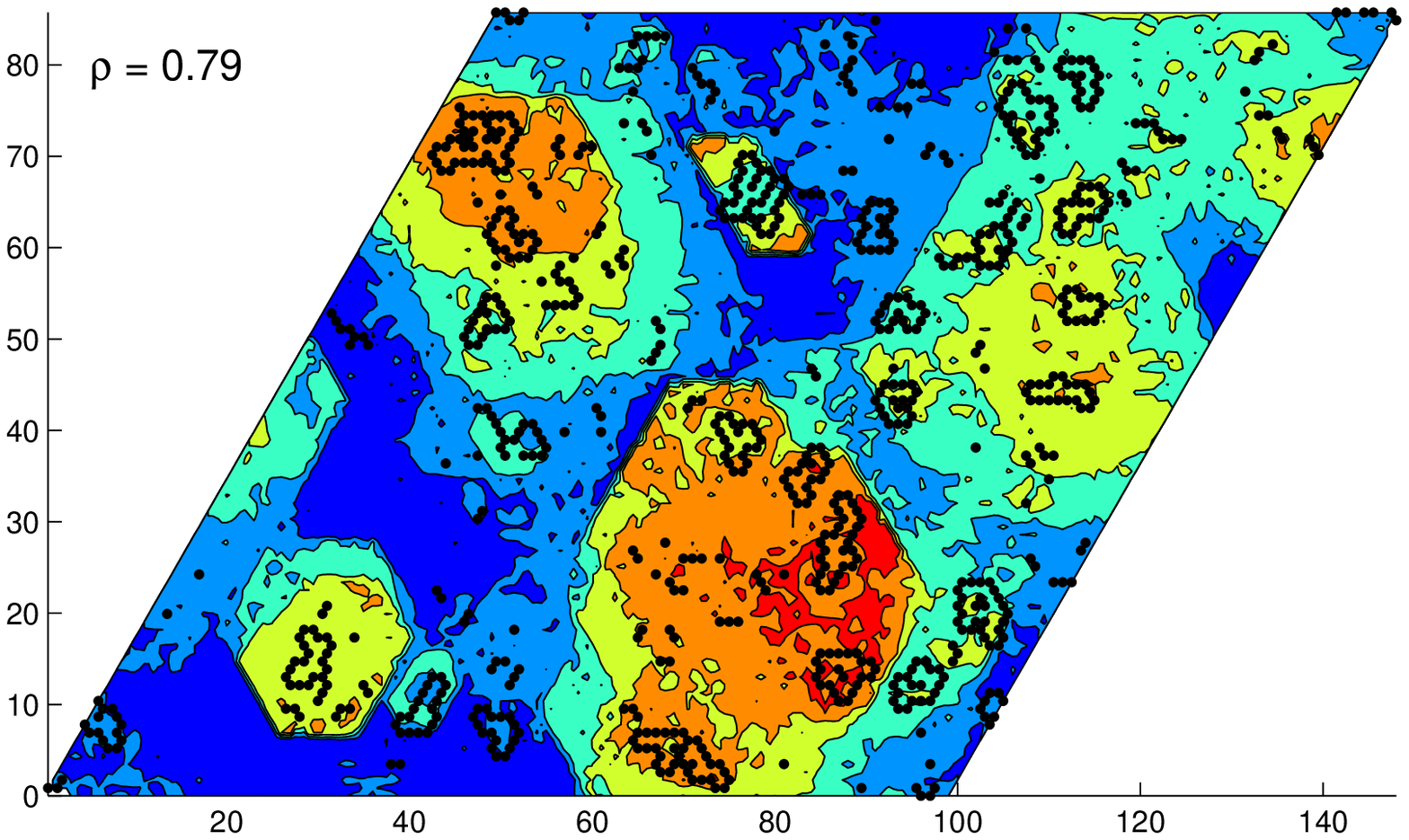}\\
    \includegraphics[width=0.5\columnwidth]{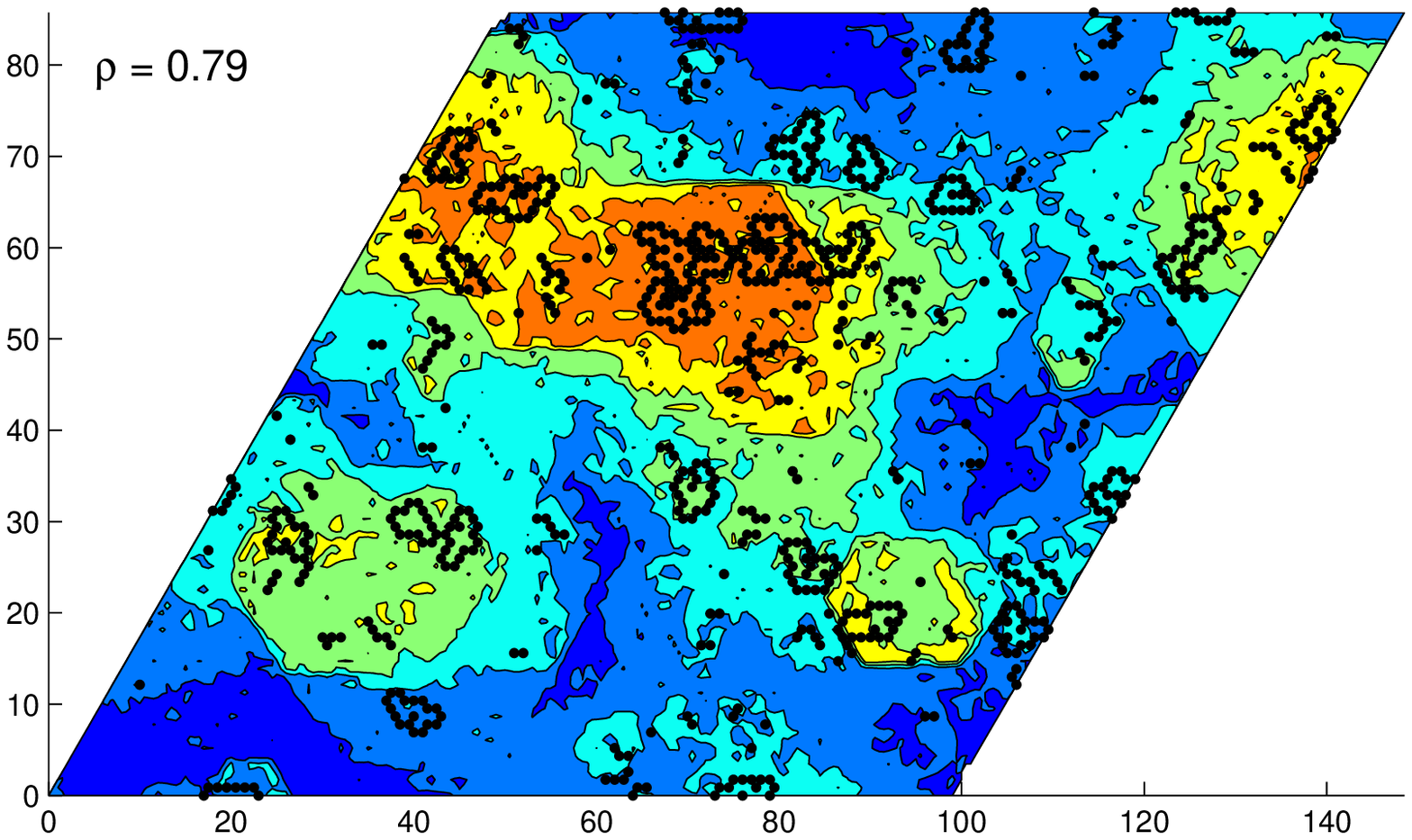}\\
    } }
\caption{Two propensity maps and clusters of particles with high
connectivity at density $\rho=0.79$.  The first map is the same as that
of Fig.\ 3(d).}\label{cluster-maps}
\end{figure}

\begin{figure}[t]
\centerline{\hbox{ \hspace{0.0in}
  \includegraphics[width=0.5\columnwidth]{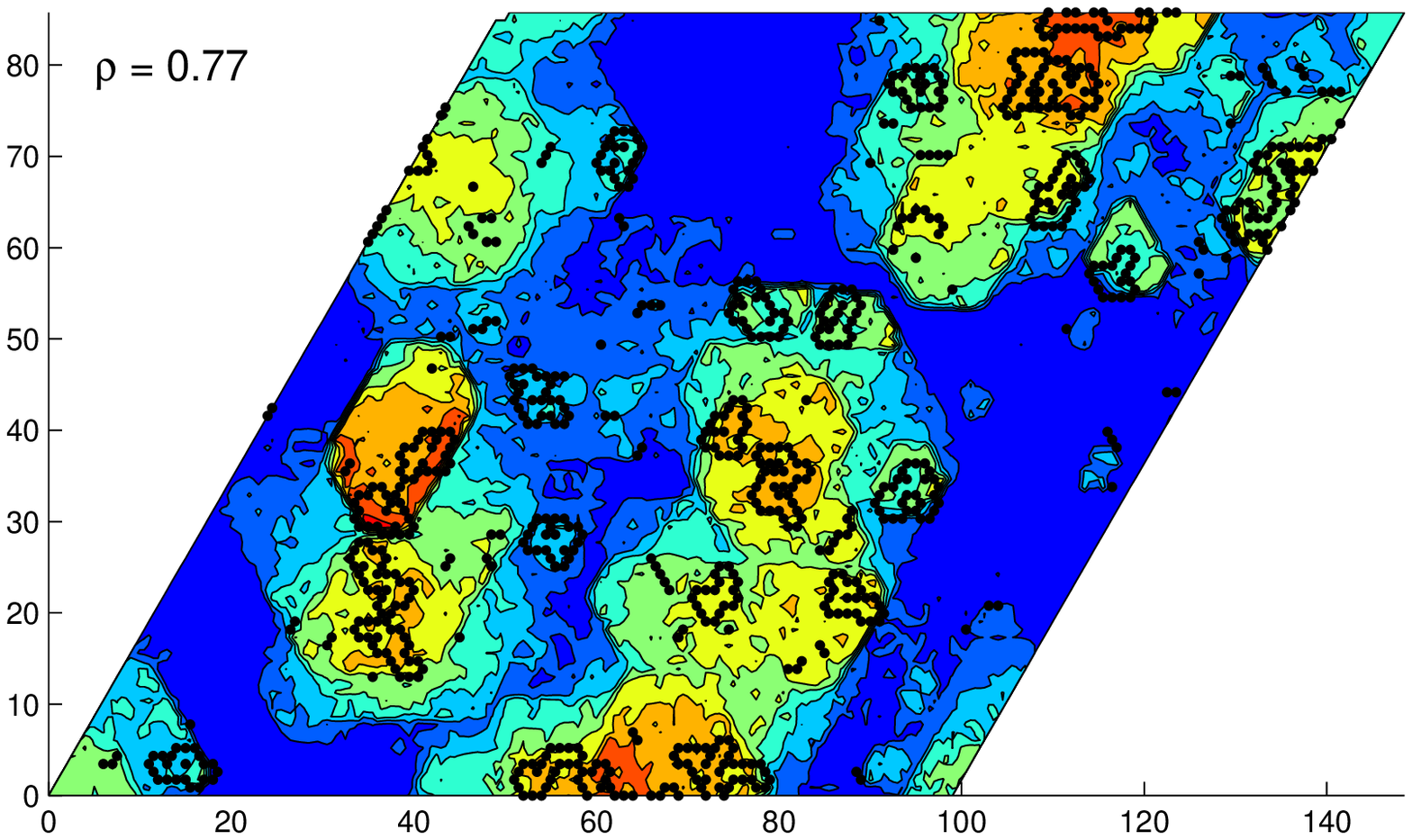}\\
  \includegraphics[width=0.5\columnwidth]{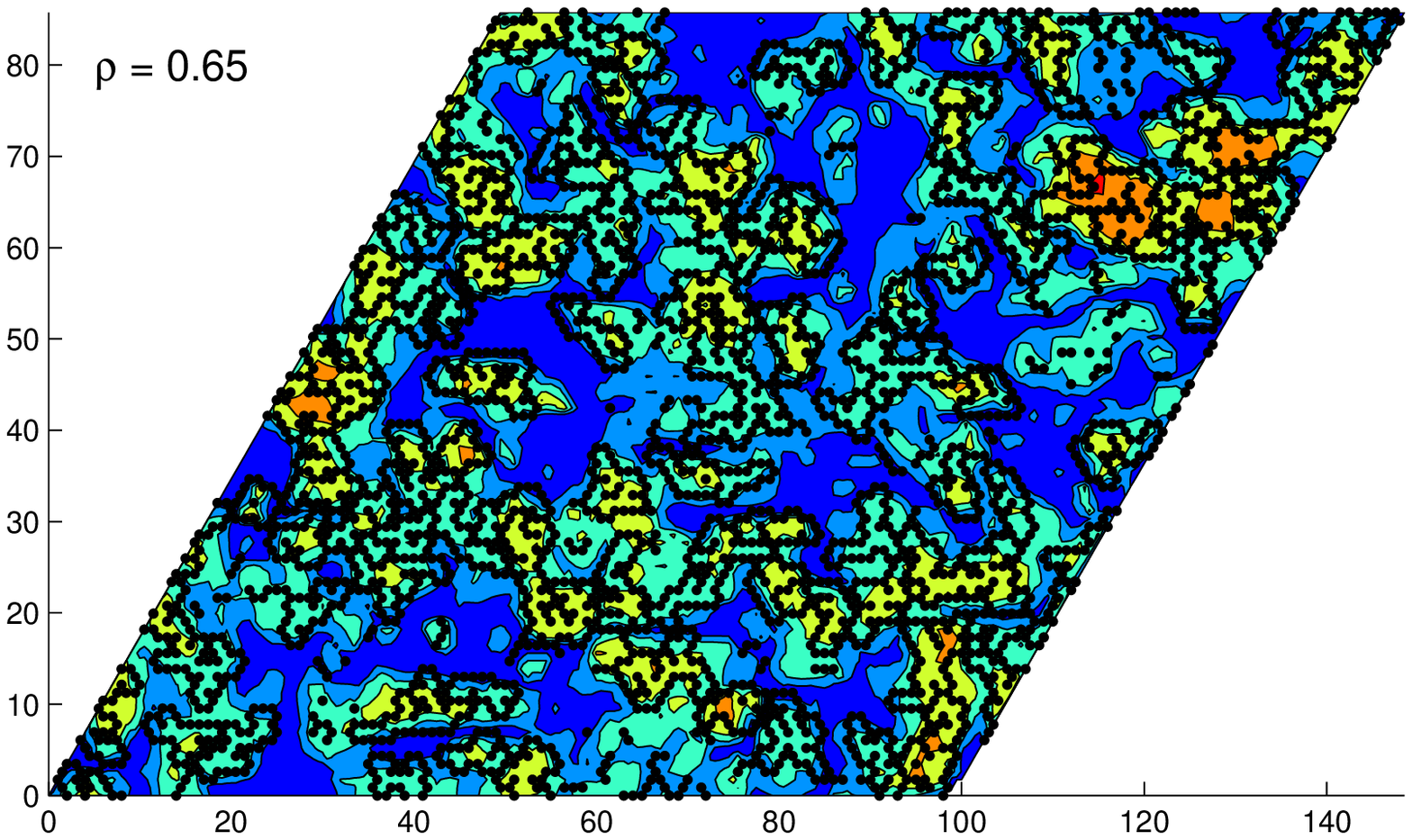}
}}
\caption{Same as previous figure, but now at two lower densities.}\label{cluster-maps-2}
\end{figure}

\begin{figure}[t]
\centering
 \includegraphics[width=0.6\columnwidth]{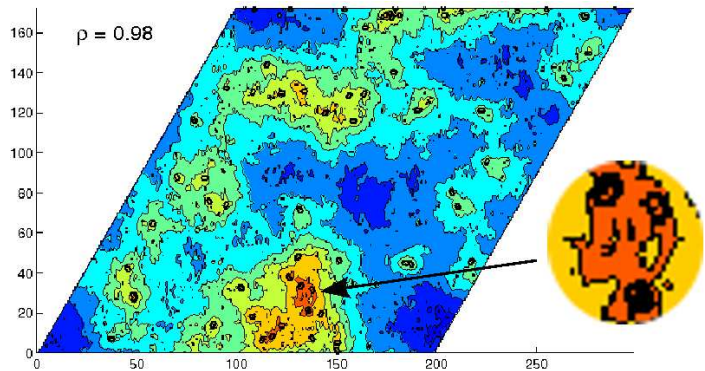}\\
\caption{Propensity map and clusters of particles with high
connectivity in the (1)-TLG at density $\rho=0.98$.}
\label{cluster-maps-1tlg}
\end{figure}

Fig.\ \ref{cluster-maps} shows two propensity maps at the highest
density we studied, $\rho=0.79$.  Highlighted in each panel (black
circles) are those particles whose cluster connectivities are above a
chosen threshold (highest 10\% in this case). These particles
typically form closed rings around a large vacancy cluster. It is
clear that the most highly connected particles correlate extremely
well with those particles showing the highest propensity.  This is not
only true for the highest density configurations, for which the
propensity distribution is most heterogeneous, a good agreement is
also observed at lower density.  In Fig.\ \ref{cluster-maps-2} we show
the same for densities $\rho=0.77$ and $\rho=0.65$ (the threshold for
$\rho=0.77$ is 10\%, and for $\rho=0.65$ is 50\%).  For the lower
density of $\rho=0.65$ the high propensity regions can almost be
entirely described by the particles whose cluster connectivity falls
into the top 50\% of the system.  

In Fig.\ \ref{cluster-maps-1tlg} we show a similar propensity map for
the (1)-TLG at a high density of particles.  While less striking, the
behaviour is analogous to that of the (2)-TLG: the propensity is
heterogeneously distributed in space, and regions of high propensity
correlate with highly connected clusters.

\begin{figure}[t]
  \centerline{\hbox{ \hspace{0.0in}
    \includegraphics[angle=-90,width=0.5\columnwidth]{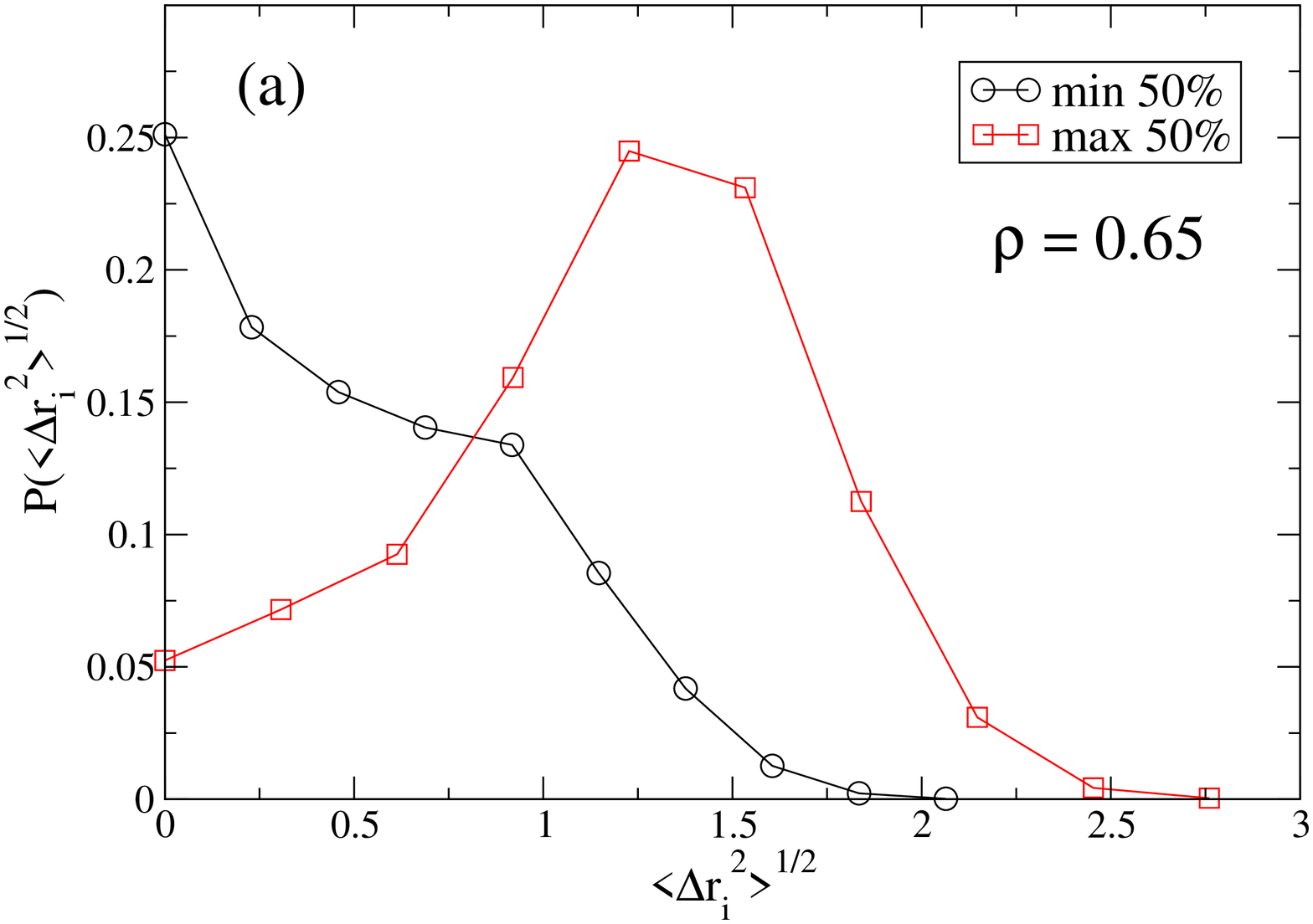}\\
    \includegraphics[angle=-90,width=0.5\columnwidth]{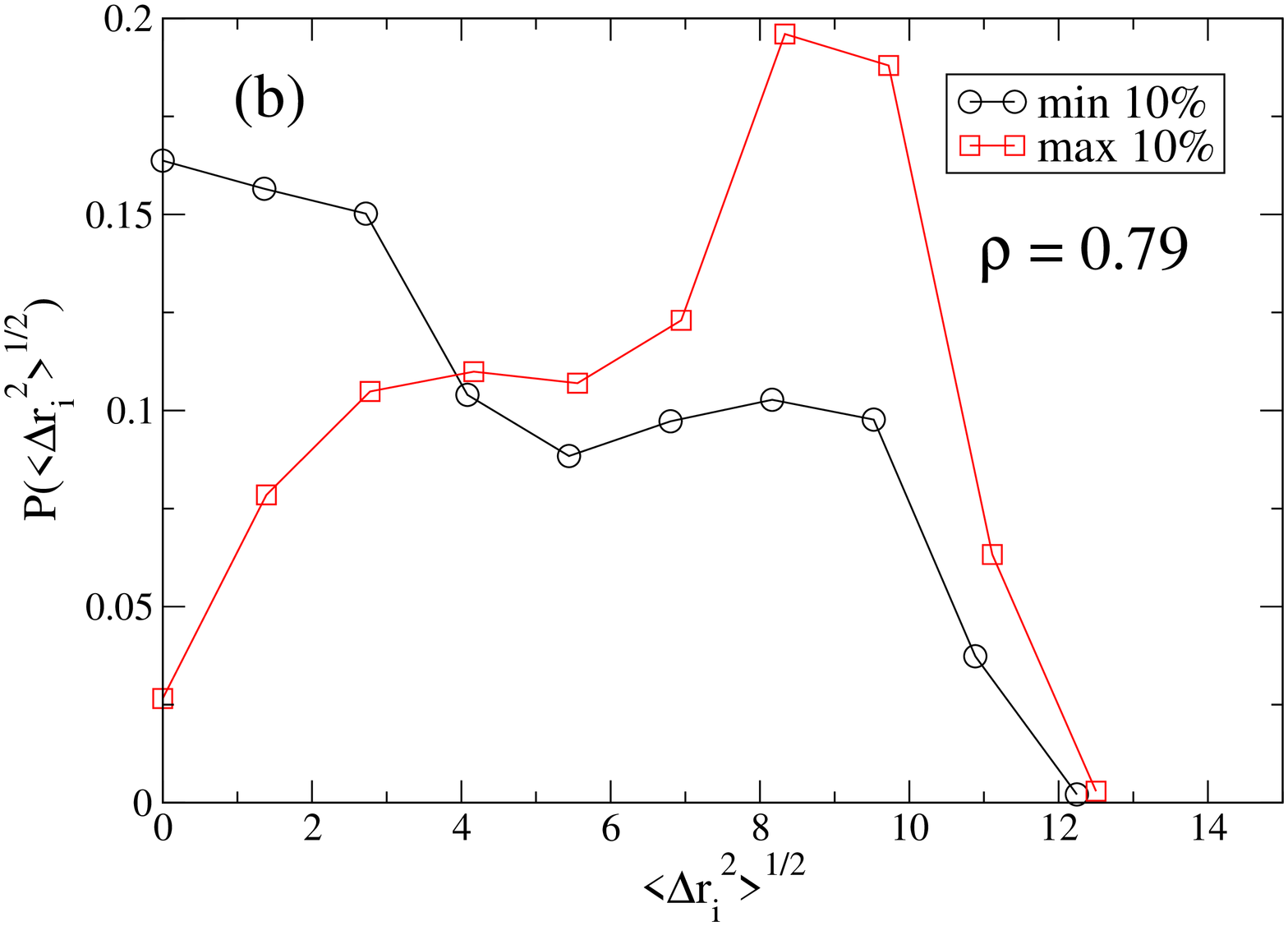}\\
    } }
    \caption{Distribution of propensity for particles which are
    connected to the largest and smallest vacancy
    clusters.}\label{propensity-clusters}
\end{figure}

The result above is also apparent when analysing the distribution of
propensity for particles with the minimum and maximum cluster
connectivity, as shown by Fig.\ \ref{propensity-clusters} for the
(2)-TLG.  In contrast to the local density measure, the two
distributions are clearly different, the particles with minimum
connectivity showing a peak at low propensity whereas the peak is
shifted to higher propensity for particles with high connectivity.

Our results here are another indication of the difficulty in trying to
connect dynamics to structure in atomistic systems.  Even in the
simple systems we consider, like the (2)-TLG and (1)-TLG, one has to
work pretty hard to obtain any significant correlation between a
structural measure and the observed dynamics, and the success has much
to do with the detailed prior knowledge of the dynamics in these
models.  Devising a similar cluster scheme which is of any use in
atomistic models appears to be a very complicated task.

\section{Time dependence of propensity}

\begin{figure}[t]
    \centering
    \includegraphics[width=0.32\columnwidth]{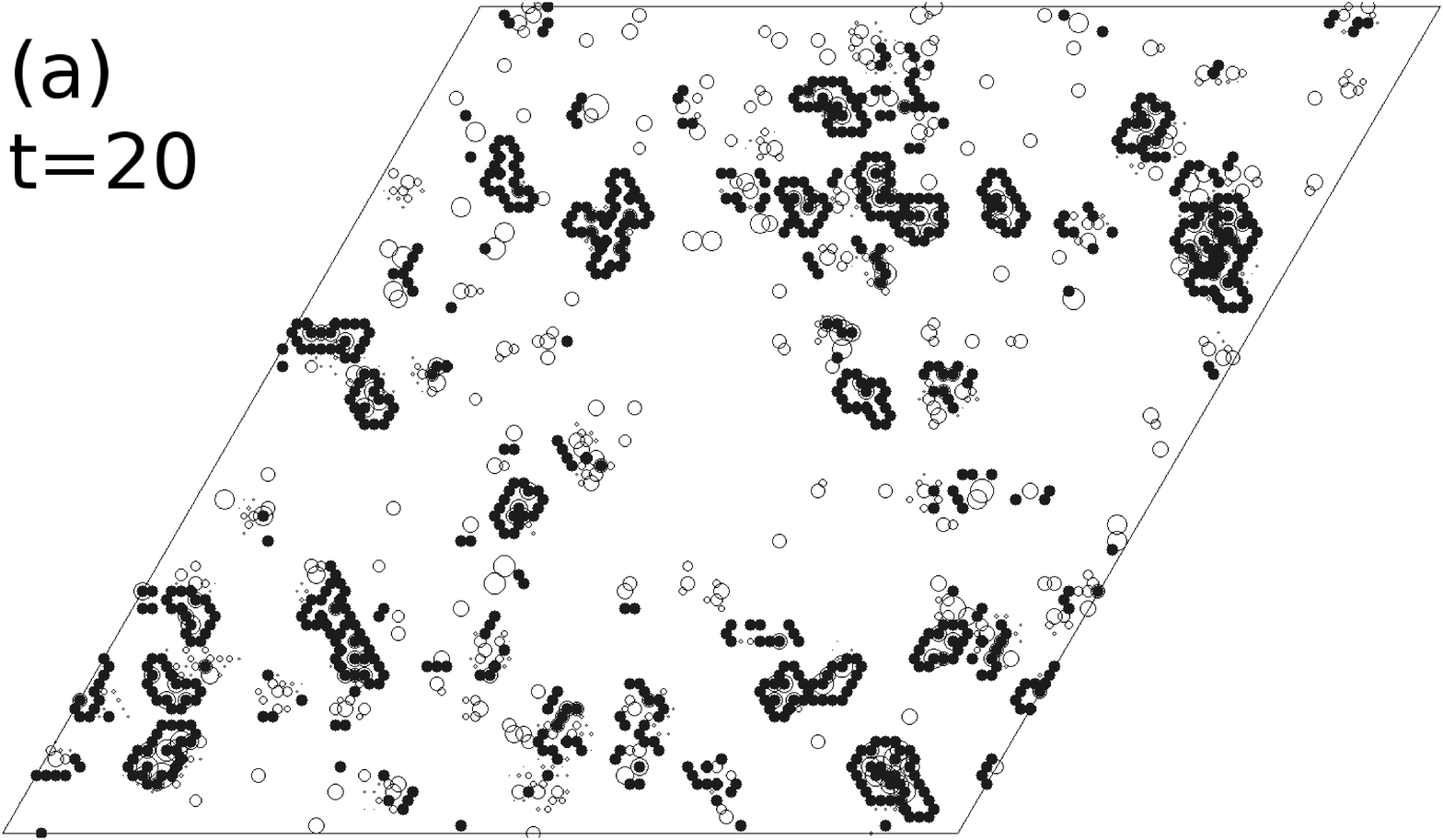}
    \includegraphics[width=0.32\columnwidth]{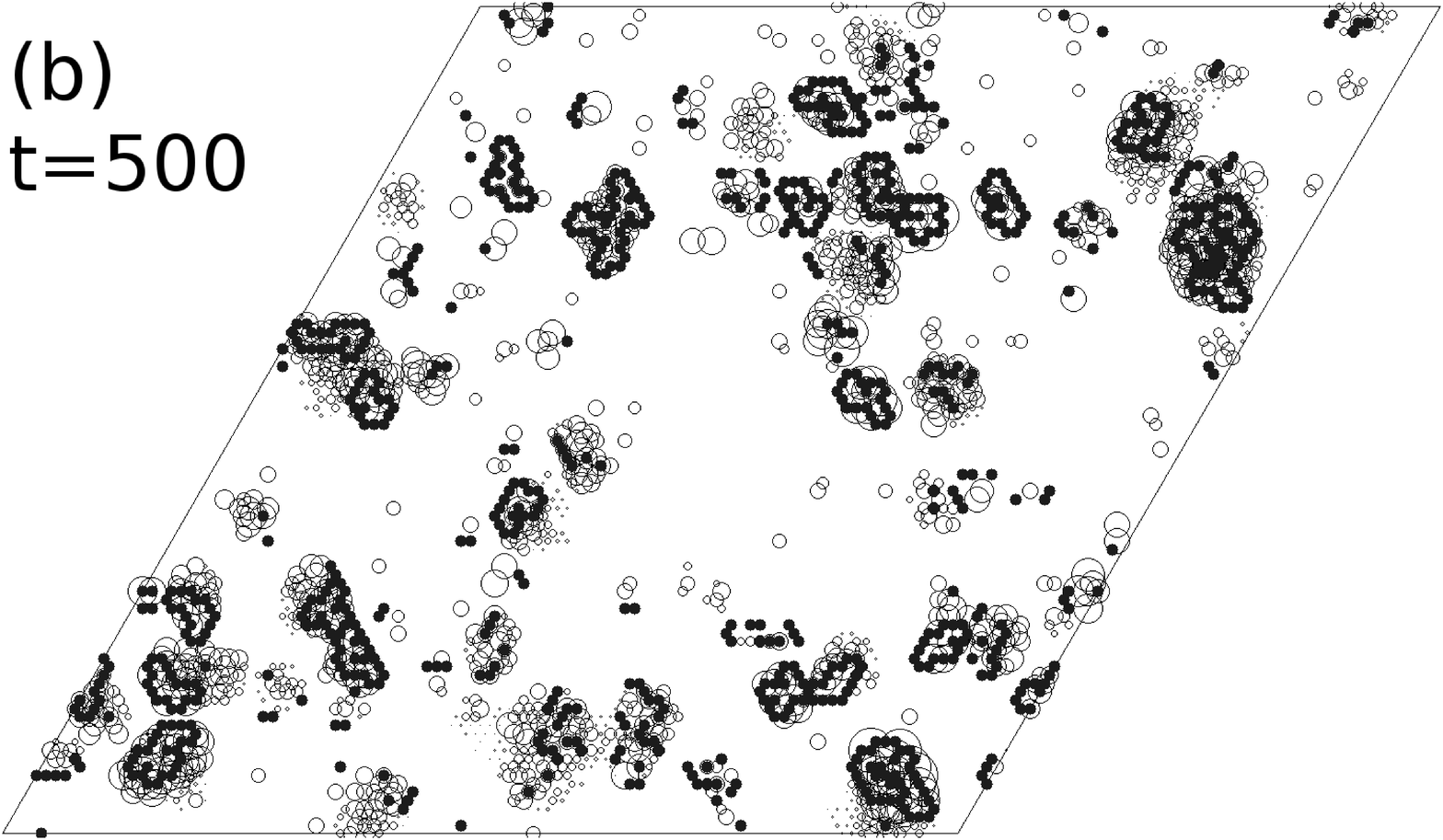}
    \includegraphics[width=0.32\columnwidth]{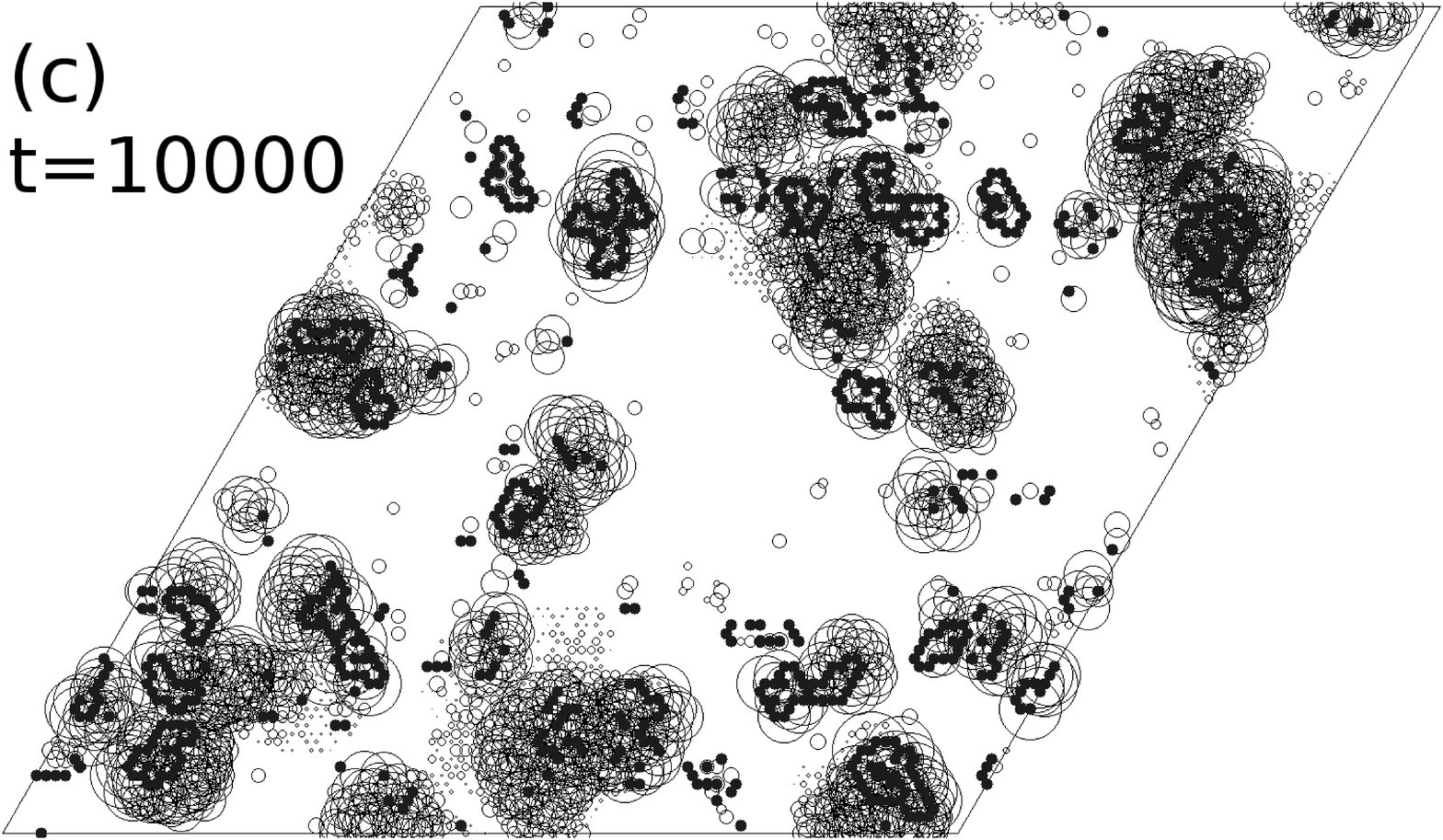}
    \caption{Time evolution of the propensity map
    \cite{Movies}.}\label{propensity-evolution}
\end{figure}

An important issue to understand is the time dependence of the
propensity.  Fig.\ \ref{propensity-evolution} shows three propensity
maps created using an initial configuration of $100\times100$ sites at
density $\rho=0.79$. The three panels show the distribution of
propensity at different time scales, (a) $t=20$, (b) $t=500$ and (c)
$t=10^4$. Here we do not use a contour plot but instead construct the
distribution from circles centered about a particle's position in the
initial configuration with a radius equal to its propensity. Filled
circles indicate the particles whose cluster connectivity is within
the top 5\% of the system.

Comparing the figures it is evident that the highly connected
particles act as seeds for mobility within the system with propensity
circles seen to emanate outwards from these hot spots.  In this sense,
propensity at shorter times is a good predictor of propensity at
longer times.  While clearly in our system there is no such thing as a
local Debye-Waller factor, this spatial correlation between short and
long time dynamics does resemble the one found in \cite{Harrowell2}.
As the system evolves some interesting features can be seen in the
propensity maps. In cases where two or more large vacancy clusters are
in close proximity one may observe the individual clusters to
coalesce. This is often accompanied by a sudden burst of motion at the
contact point as indicated by a region of newly mobile small
circles. The propensity of these particles may then rapidly grow as
motion persists in the region.  This behaviour is consistent with the
shape of the propensity distributions shown in Fig.\
\ref{propensity-distribution}.

Examining the evolution of the propensity field sheds light on the
nature of the (2)-TLG dynamics. As suggested earlier it is the
clusters of vacancies that help facilitate the cooperative
rearrangement of particles within the system. From this we find highly
connected particles which themselves form clusters that move, coalescing 
and fragmenting into clusters of different shape and size. It appears that 
these clusters underly the dynamics and are in effect the essential excitations 
for the model. The initial configuration represents only a single
instantaneous indication of the system's structure. By labeling
particles according to their cluster connectivity we only manage to
capture the hot spots at a particular moment in time. By analysing the
evolution of the most highly connected particles we should be able to
track the motion of the excitations throughout the lattice.  Note this
is not the same as simply looking at the motion of vacancies.

\begin{figure}[b]
    \centering
    \includegraphics[width=0.32\columnwidth]{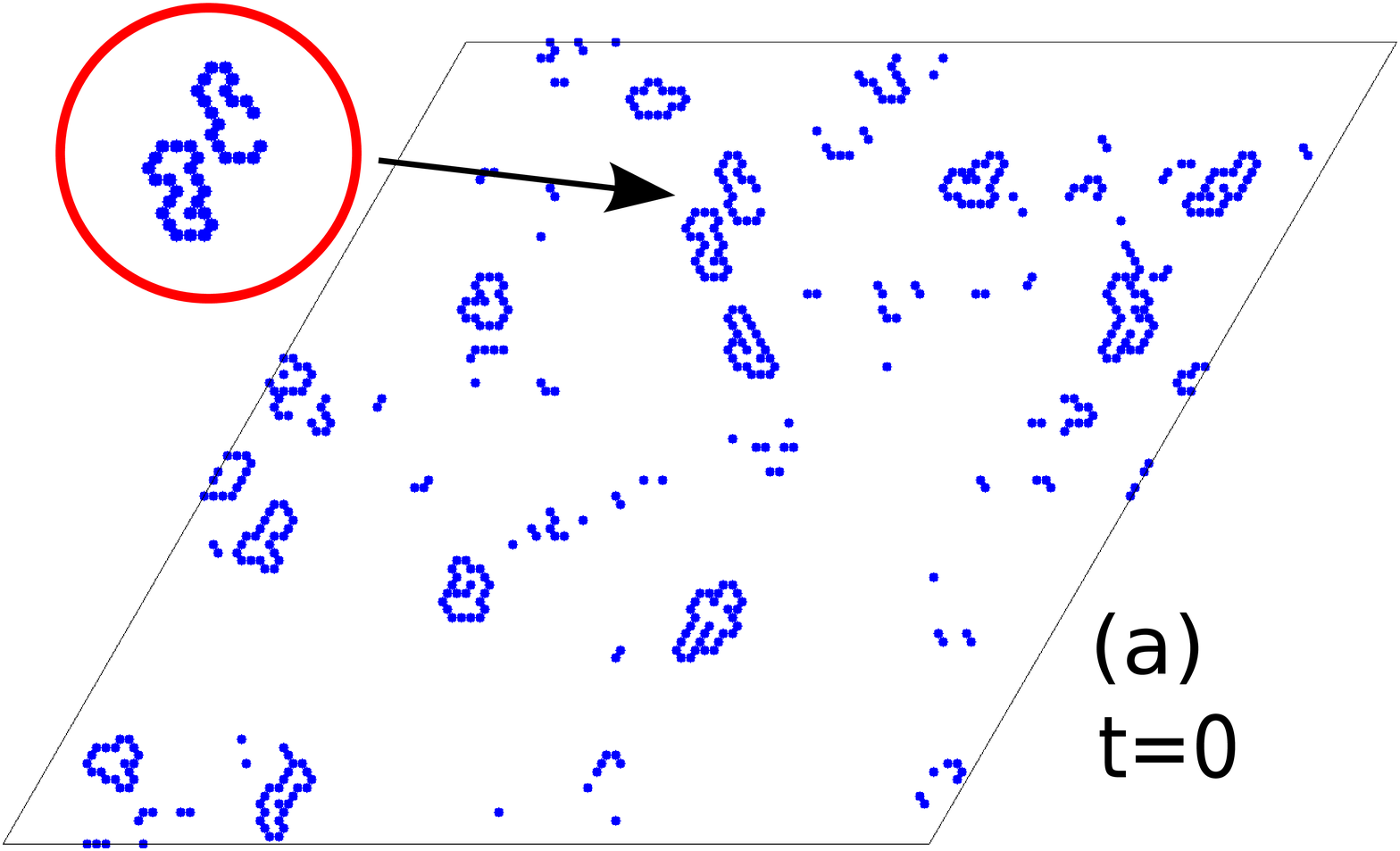}
    \includegraphics[width=0.32\columnwidth]{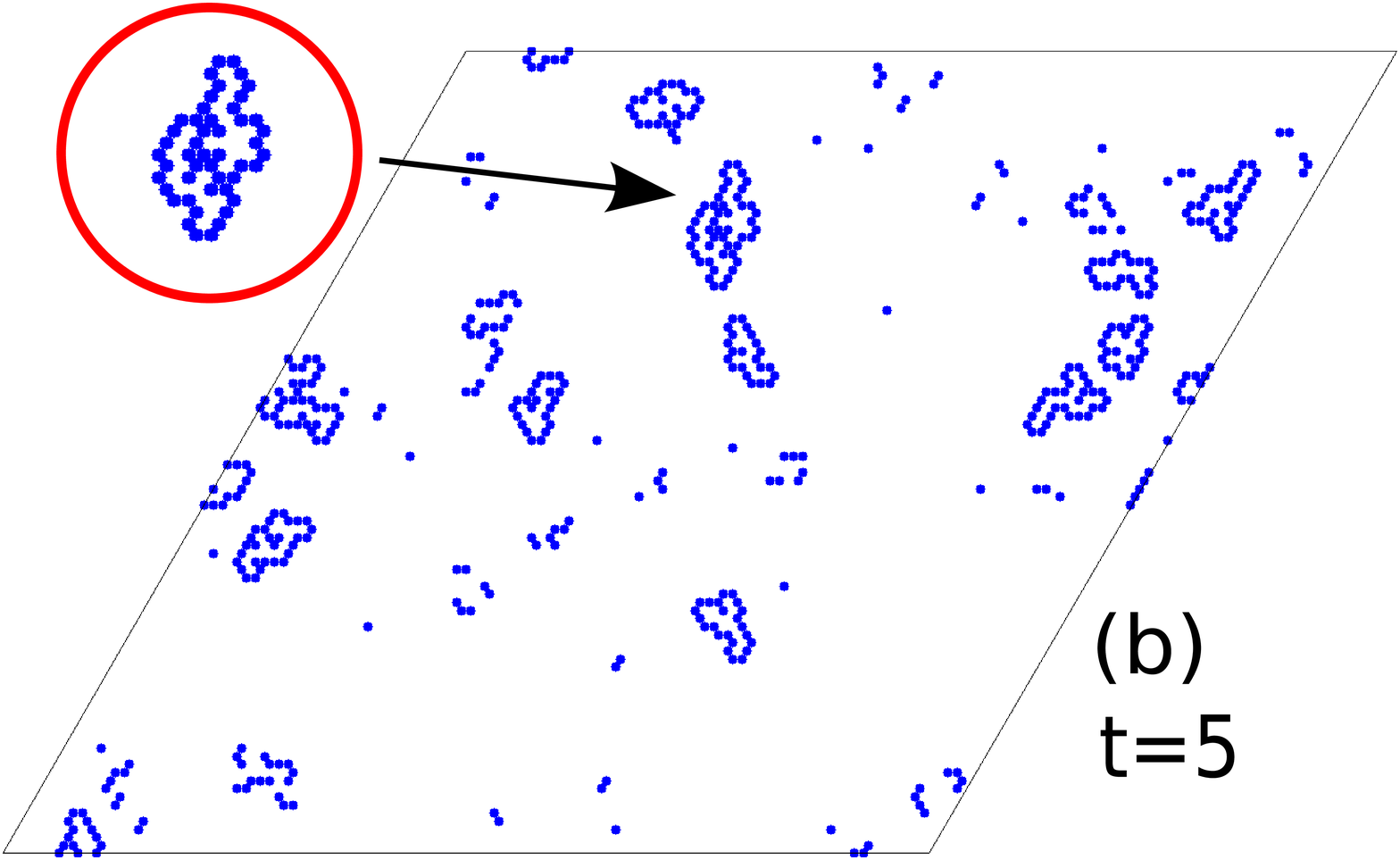}
    \includegraphics[width=0.32\columnwidth]{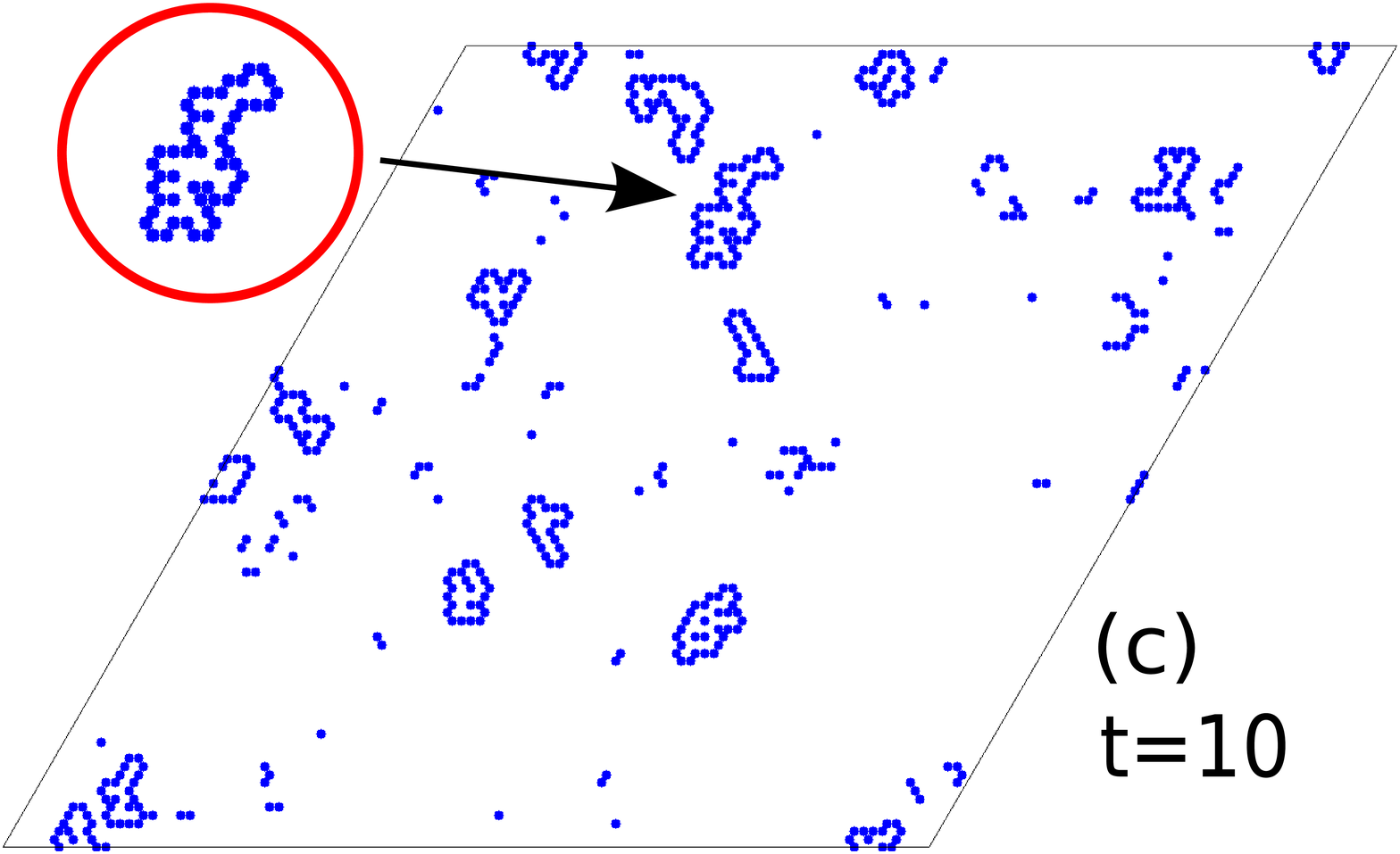}
    \caption{The motion of particles connected to the largest vacancy
      clusters \cite{Movies}.}\label{cluster-motion}
\end{figure}

The motion of one of the clusters is illustrated in Fig.\
\ref{cluster-motion}.  Here we show the position of the particles with
the top 5\% connectivity at three different times. Highlighted in the
three panels is the motion of two nearby clusters, which are seen to
come together and eventually coalesce.  Fig.\ \ref{cluster-motion}
shows two characteristic lengthscales, the typical cluster size and
the typical distance between them.  This resembles the super-defect
scenario of \cite{TBF}.  One difference however, at least
superficially, appears to be the dynamics: super-defects are
supposed to be free to propagate; the clusters we find have pinned
end-points, they fluctuate and may combine to form larger clusters,
but are not free to move away.  In any case, it would be worth
exploring the connection to super-defects further.

\section{Propensity correlations and time-correlations}

We end by considering time correlations and
cross-correlations or overlaps within the propensity ensemble.  The
normalized autocorrelation function is $C(t) \equiv \langle n_i(0)
n_i(t) \rangle_{\rm c} / \langle n_i(0) n_i(0) \rangle_{\rm c}$, where
$n_i=\{0;1\}$ specifies the site occupancy, and $\langle \cdot
\rangle_{\rm c}$ indicates connected part of correlations, $\langle n
n' \rangle_{\rm c} \equiv \langle n n' \rangle - \rho^2$.  For large
enough systems $C(t)$ calculated in the propensity ensemble coincides
with the standard autocorrelation function.  A second interesting
function is the cross-correlation or overlap $Q(t)$ between two
different configurations at time $t$ in the propensity ensemble, $Q(t)
\equiv \langle n_{i,\alpha}(t) n_{i,\beta}(t) \rangle_{\rm c} /
\langle n_{i,\alpha}(0) n_{i,\beta}(0) \rangle_{\rm c}$, where
$\alpha,\beta$ indicate different trajectories with identical initial
conditions.

\begin{figure}[t]
    \centering
    \includegraphics[angle=0,width=0.32\columnwidth]{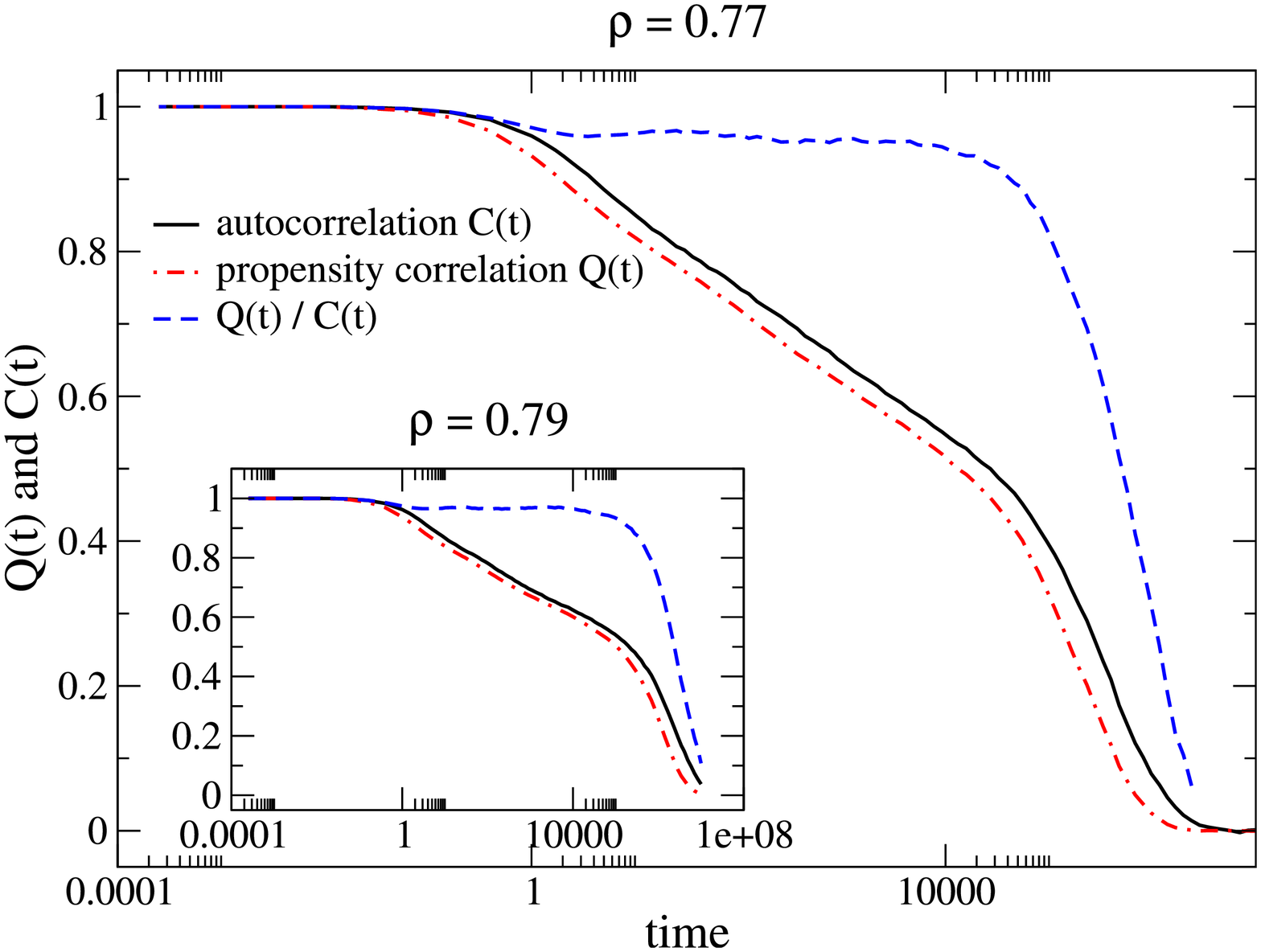}
    \includegraphics[angle=0,width=0.32\columnwidth]{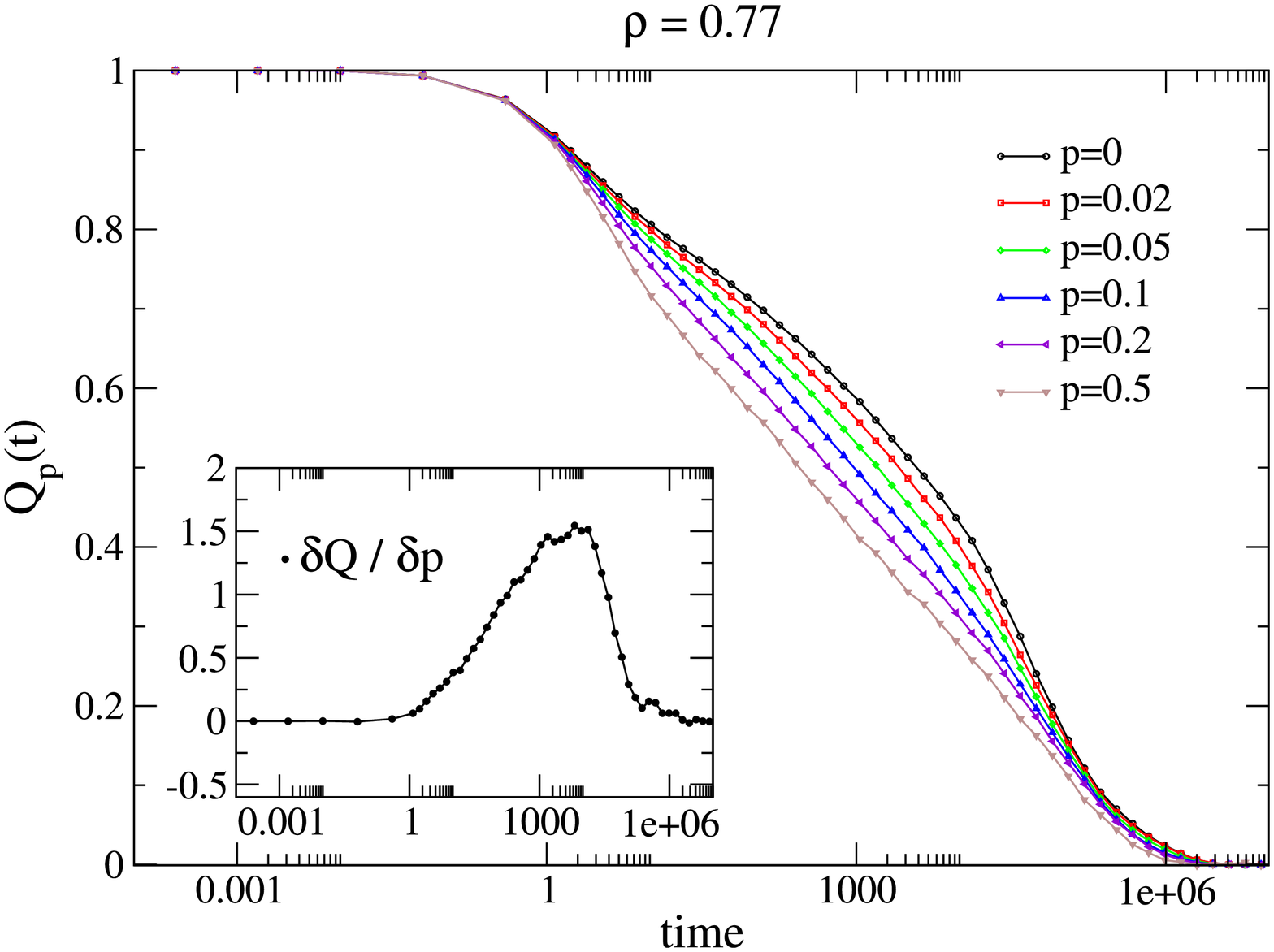}
    \includegraphics[angle=0,width=0.32\columnwidth]{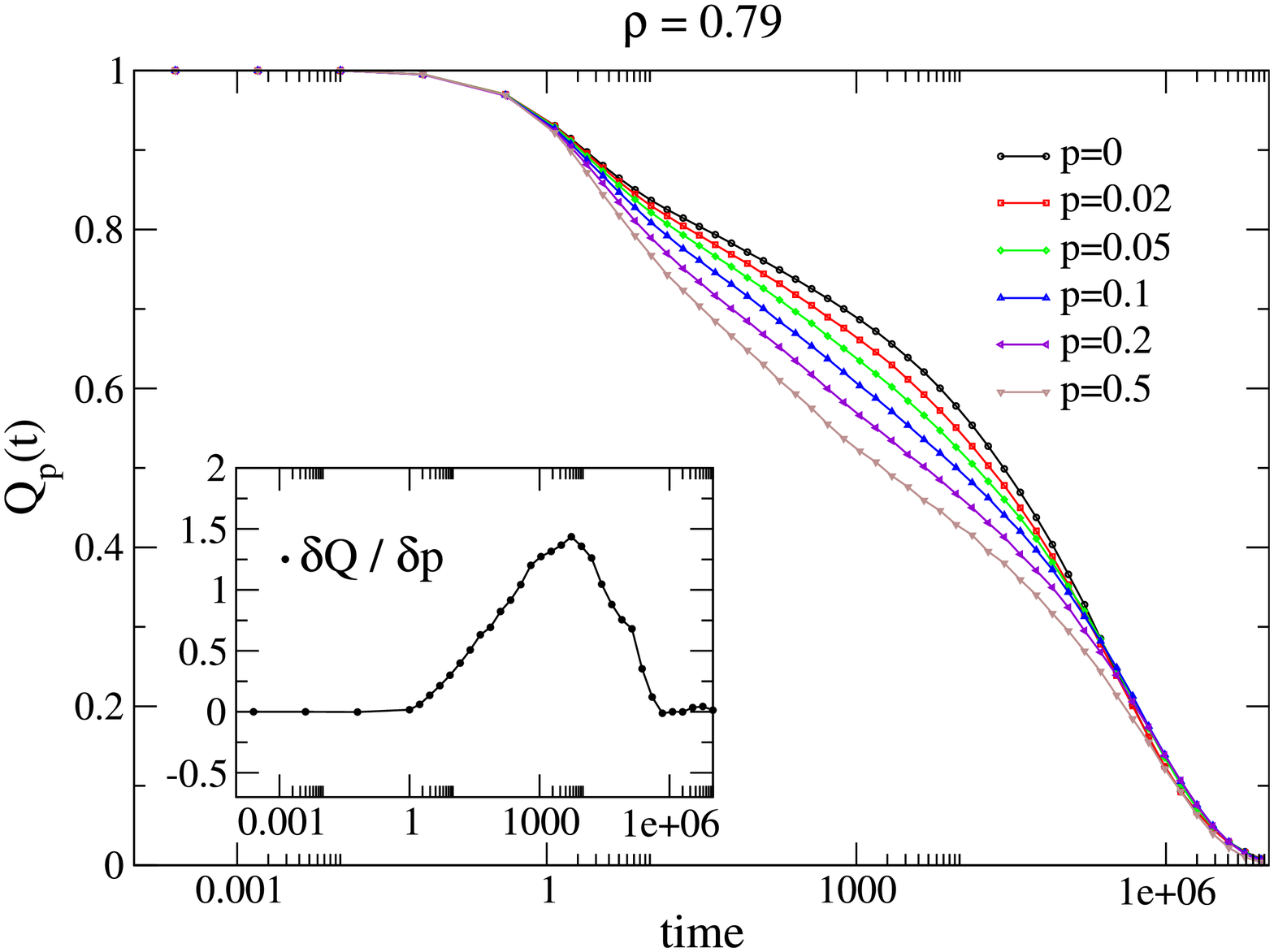}
    \caption{(a) Autocorrelation function $C(t)$ and propensity
    overlap $Q(t)$ in the (2)-TLG for two different densities $\rho$
    (main panel and inset).  The dashed line is the ratio $Q(t)/C(t)$.
    (b) Overlaps $Q_p(t)$ for different values of $p$ at density
    $\rho=0.77$.  The inset shows the response $\left. \delta
    Q_p(t)/\delta p \right|_{p=0}$.  (c) Same as (b) but for density
    $\rho=0.79$. }\label{Q}
\end{figure}

Figure \ref{Q}(a) shows $C(t)$ and $Q(t)$ for the (2)-TLG.  As
expected, $C(t) \geq Q(t)$ for all $t$, i.e., the distance between
initial and final configurations along a trajectory is always larger
than that between two different final states.  In fact, for any
reversible dynamics $Q(t)=C(2t)$ as long as the initial conditions are
in equilibrium (for a discussion of the behaviour of $Q$ in aging see
e.g. \cite{Barrat}).  Interestingly, when $C(t)$ is much slower than
an exponential, as in the present case where it appears to be
logarithmic, the ratio $Q(t)/C(t)$ remains almost constant at long
times.  This is to be contrasted to the simple non-interacting lattice
gas, where all decorrelation is through the diffusion of vacancies,
and for which $Q(t)/C(t) \approx C(t)$.

The above illustrates how trajectories within the propensity ensemble
are correlated through their common initial conditions.  The
sensitivity to the initial state can be tested by the response to
perturbations in the initial conditions.  Consider the case of two
trajectories $\alpha$ and $\beta$ whose starting configurations differ
only in the location of a fraction $p$ of their vacancies.  We denote
by $Q_p(t)$ their normalized overlap.  In the case of a
non-constrained lattice gas we have $Q_p(t)=Q_{p=0}(t)$, i.e., the
initial distance is preserved in the trajectories at all times.
Figs.\ \ref{Q}(b) and (c) show the overlaps for the (2)-TLG.  In this
case the distance between configurations increases.  The corresponding
response, $\left. \delta Q_p(t)/\delta p \right|_{p=0}$, shown in the
insets, is non-monotonic and peaks at times around the relaxation
time.  

As argued in \cite{Mauro,Rob}, sensitivity to initial conditions
should be a signature of the dynamic phase transition which underlies
dynamic heterogeneity in facilitated models in general: dynamics takes
place very close to phase coexistence between an active phase, where
motion is plentiful, and an inactive phase, where motion is scarce.
Due to this phase equilibrium, initial conditions play an important
role since they act as boundary fields in space-time.  Dynamic
propensity appears to be a useful tool to study their relevance.

\ack

We thank Peter Harrowell for correspondence.  This work was supported
by EPSRC grants GR/R83712/01, GR/S54074/01, and University of
Nottingham grant FEF3024.

\end{document}